\documentclass{article}

\usepackage{arxiv}

\usepackage[utf8]{inputenc} 
\usepackage[T1]{fontenc}    
\usepackage{hyperref}       
\usepackage{url}            
\usepackage{booktabs}       
\usepackage{longtable} 
\usepackage{amsfonts}       
\usepackage{nicefrac}       
\usepackage{amsmath, amssymb, amsthm}
\usepackage{microtype}      
\usepackage{lipsum}		
\usepackage{graphicx}
\usepackage{natbib}
\usepackage{doi}
\usepackage{lscape} 
\usepackage{hyperref}
\usepackage{cleveref}
\usepackage[normalem]{ulem}
\useunder{\uline}{\ul}{}
\usepackage[table]{xcolor} 
\usepackage{booktabs} 
\usepackage{caption} 
\usepackage{bm}
\usepackage{mathtools}
\usepackage{bbm}
\usepackage{dsfont}
\usepackage{multicol}
\usepackage{multirow}

\newtheorem{theorem}{Theorem}
\renewcommand{\P}{\mathrm{P}}
\newtheorem{assumption}{Assumption}
\DeclareMathOperator*{\argmin}{arg\min}

\title{Penalized FCI for Causal Structure Learning in a Sparse DAG for Biomarker Discovery in Parkinson's Disease}

\author{Samhita Pal\thanks{
    spal4@ncsu.edu}\hspace{.2cm}\\
    Department of Statistics, North Carolina State University\\
    Dhrubajyoti Ghosh \\
    Department of of Biostatistics and Bioinformatics, Duke University\\
    and \\
    Shu Yang\\
    Department of Statistics, North Carolina State University}

\renewcommand{\shorttitle}{\textit{arXiv} Template}

\hypersetup{
pdftitle={A template for the arxiv style},
pdfsubject={q-bio.NC, q-bio.QM},
pdfauthor={David S.~Hippocampus, Elias D.~Striatum},
pdfkeywords={First keyword, Second keyword, More},
}

\begin{document}
\maketitle

\begin{abstract}
Parkinson’s disease (PD) is a progressive neurodegenerative disorder that lacks reliable early-stage biomarkers for diagnosis, prognosis, and therapeutic monitoring. While cerebrospinal fluid (CSF) biomarkers, such as $\alpha$-synuclein seed amplification assays ($\alpha$Syn-SAA), offer diagnostic potential, their clinical utility is limited by invasiveness and incomplete specificity. Plasma biomarkers provide a minimally invasive alternative, but their mechanistic role in PD remains unclear. A major challenge is distinguishing whether plasma biomarkers causally reflect primary neurodegenerative processes or are downstream consequences of disease progression. To address this, we leverage the Parkinson’s Progression Markers Initiative (PPMI) Project 9000, containing 2,924 plasma and CSF biomarkers, to systematically infer causal relationships with disease status. However, only a sparse subset of these biomarkers and their interconnections are actually relevant for the disease. Existing causal discovery algorithms, such as Fast Causal Inference (FCI) and its variants, struggle with the high dimensionality of biomarker datasets under sparsity, limiting their scalability. We propose Penalized Fast Causal Inference (PFCI), a novel approach that incorporates sparsity constraints to efficiently infer causal structures in large-scale biological datasets. By applying PFCI to PPMI data, we aim to identify biomarkers that are causally linked to PD pathology, enabling early diagnosis and patient stratification. Our findings will facilitate biomarker-driven clinical trials and contribute to the development of neuroprotective therapies.
\end{abstract}


\section{Introduction}
\label{sec:intro}
Parkinson’s disease (PD) is a progressive neurodegenerative disorder that begins at the molecular and cellular level long before the onset of motor symptoms. Despite its increasing prevalence and the profound impact on patient care, there are no early-stage molecular biomarkers for diagnosis, prognosis prediction, or monitoring therapeutic response. This absence of reliable biomarkers significantly hinders translational research and patient management, evidenced by the fact that Levodopa remains the standard of care more than 50 years after its introduction. PD progresses from a pre-motor stage, characterized by non-motor symptoms such as REM sleep behavior disorder, to the disabling motor stage. Identifying objective biomarkers for these early and pre-motor stages is crucial to enable early intervention and curb the underlying neurodegenerative process. However, current biomarker research faces major limitations. One promising approach is the use of $\alpha$-synuclein seed amplification assays ($\alpha$Syn-SAA) \citep{shahnawaz2017development}, which can identify misfolded $\alpha$-synuclein in cerebrospinal fluid (CSF) and serve as a proxy for Lewy body pathology \citep{braak2003staging, simuni2018baseline}. While this represents a significant advance, $\alpha$Syn-SAA has notable drawbacks: it requires an invasive lumbar puncture, lacks information about disease severity or progression, and has limited utility in patients with LRRK2 mutations or those with nigrostriatal neurodegeneration without Lewy body pathology \citep{siderowf2023assessment}. These limitations highlight the pressing need for additional biomarkers that are both minimally invasive and capable of reflecting early neurodegenerative processes in PD.

Given the limitations of CSF biomarkers, there is an urgent demand for peripheral fluid biomarkers that can be used for large-scale, repeated, and long-term monitoring. Blood-based biomarkers, particularly those detectable in plasma, offer a less invasive and more accessible alternative for disease tracking and clinical trials. In Alzheimer’s disease research, the development of blood-based biomarkers has transformed the field, enabling early diagnosis, monitoring of disease progression, and even patient stratification for novel therapeutic interventions \citep{budd2022two, van2023lecanemab, sims2023donanemab}. PD research must follow suit. The Parkinson’s Progression Markers Initiative (PPMI) provides an unparalleled opportunity to discover and validate plasma biomarkers. As a well-structured, longitudinal study, PPMI has already provided critical insights into PD pathophysiology, and now, it must expand efforts into identifying blood-based biomarkers that can detect pre-symptomatic neurodegeneration. While neurofilament light chain (NfL) has emerged as a promising serum biomarker correlating with motor and cognitive progression, it is not specific to PD \citep{mollenhauer2019validation}. We need additional biomarkers that are directly tied to PD-specific neurodegenerative mechanisms. Alongside plasma, the discovery of new CSF biomarkers remains valuable, particularly for understanding the molecular underpinnings of PD and refining diagnostic tools. However, for widespread clinical adoption, a paradigm shift toward blood-based biomarker discovery is essential. Moreover, current technologies have made measuring a large number of biomarkers easy, thereby introducing many irrelevant biomarkers. 

In reality, only a few of the observed biomarkers causally influence the disease.The edges are also usually sparse as biological networks are modular, with biomarkers interacting within localized pathways rather than forming dense, fully connected graphs. Hence, a sparsity assumption in both edges and nodes is essential in such high-dimensional studies, as it not only provides sound statistical ground for avoiding overfitting and identifiability issues, it also helps in interpretability of the causal findings. The identification of plasma biomarkers in PPMI would facilitate earlier diagnosis, improve patient stratification for clinical trials, and support the development of neuroprotective therapies.

Recent advances in proteomic technologies have enabled large-scale analyses of CSF and plasma biomarkers for PD, offering new insights into disease pathophysiology and potential early diagnostic markers. Several studies have identified promising plasma-based biomarkers, including apolipoprotein A1, which is associated with age at PD onset and dopaminergic integrity \citep{qiang2013plasma}, as well as panels of inflammatory and neurodegeneration-related proteins, such as interleukin-6, cystatin B, fibroblast growth factor 21, and peptidase inhibitor 3, which correlate with cognitive decline and motor symptom progression \citep{bartl2023blood}. Additionally, machine learning models leveraging plasma biomarkers, including granulin precursor, complement C3, and prostaglandin-H2-D-isomerase, have successfully identified pre-motor individuals up to seven years before symptom onset \citep{hallqvist2024plasma}. CSF biomarkers, such as DOPA decarboxylase (DDC), have shown strong diagnostic and prognostic potential, being consistently upregulated across multiple cohorts and correlating with clinical severity \citep{rutledge2024comprehensive}. CSF and blood biomarkers, including $\alpha$-synuclein species, lysosomal enzymes, amyloid and tau markers, and neurofilament light chain, have shown promise for improving PD diagnosis and prognosis, though further validation in large cohorts is needed \citep{parnetti2019csf}. While plasma biomarkers present a non-invasive alternative to CSF markers, their mechanistic specificity remains a challenge, as many are systemic proteins whose relationship to neurodegeneration is not fully understood. Current studies predominantly rely on association-based analyses, leaving open the question of causality in biomarker relationships. This gap underscores the need for causal discovery approaches to disentangle direct from indirect biomarker effects and enhance the clinical utility of plasma biomarkers for early diagnosis, prognosis, and patient stratification in PD.

A causal discovery framework using a Directed Acyclic Graph (DAG) offers a principled approach to uncovering the interplay between CSF and plasma biomarkers. Unlike traditional correlation-based analyses, DAG-based methods can infer directional relationships, distinguishing whether plasma biomarkers reflect downstream consequences of neurodegeneration or if they can serve as early indicators of disease progression. This is particularly important for identifying minimally invasive plasma biomarkers that reliably track PD pathology. By applying causal discovery techniques to PPMI data, we aim to differentiate primary from secondary biomarkers, ensuring that plasma-based markers are not merely passive correlates of CSF changes, and enhance biomarker-driven clinical trials by prioritizing biomarkers that are causally linked to disease mechanisms rather than confounded by external factors.

Graphical models, particularly DAGs, have been widely studied in various fields for their ability to represent causal relationships among variables. Traditional DAG models are often used to understand causal structures in high-dimensional data, with applications in genomics, neuroscience, and social networks. Causal discovery algorithms are designed to infer the structure of a DAG that represents the causal relationships among a set of variables. These algorithms typically fall into two broad categories: constraint-based and score-based approaches. The constraint-based methods, such as the PC \citep{kalisch2007estimating} and Fast Causal Inference (FCI) \citep{spirtes2001causation,spirtes2001anytime} algorithms, rely on conditional independence tests to uncover relationships between variables. The PC algorithm is widely used when there are no unobserved confounders, identifying the skeleton of the graph and orienting edges based on test results. However, in more complex settings involving latent variables and selection bias, the FCI algorithm extends PC by detecting hidden confounders and producing a partial ancestral graph (PAG), which accounts for latent variables while maintaining some ambiguity in edge orientations. Really Fast Causal Inference (RFCI) \citep{colombo2012learning}, a variant of FCI, accelerates the discovery process by simplifying independence tests, making it more feasible for large datasets. Score-based methods, such as the Greedy Equivalence Search (GES), evaluate possible graph structures based on their fit to the data. These algorithms navigate the space of potential DAGs, seeking the highest-scoring graph using statistical likelihood functions. Such methods are flexible but computationally intensive, and they often assume the absence of latent confounding unless explicitly modified. Hybrid methods such as Greedy Fast Causal Inference (GFCI) \citep{ogarrio2016hybrid}, combine constraint-based procedures with score-based techniques such as scoring functions, like Bayesian Information Criterion (BIC), in order to improve accuracy in latent variable settings.

Many causal discovery algorithms, such as FCI, are computationally infeasible for large-scale biomarker networks due to their dependence on graph complexity, while alternatives like RFCI and GFCI, though faster, still scale poorly in high-dimensional settings. A common assumption in causal inference is that the true underlying DAG is sparse, meaning that while the number of biomarkers (nodes) may be large, only a subset exhibit direct causal relationships. However, existing methods do not fully leverage this sparsity assumption in an efficient manner, limiting their applicability to large-scale biological datasets. Our proposed Penalized Fast Causal Inference (PFCI) method provides an efficient and scalable solution for causal biomarker discovery in PPMI data, addressing key challenges that existing methods struggle to resolve. To overcome these challenges, PFCI employs a two-step approach that efficiently extracts causal relationships from high-dimensional biomarker data. First, we use penalized neighborhood selection \citep{meinshausen2006high} to construct an initial undirected dependency graph, enforcing sparsity constraints to retain only the most relevant connections before applying FCI for causal orientation. This significantly reduces the computational burden while maintaining accuracy. The second step applies FCI on this reduced structure, orienting edges while accounting for possible latent confounders, a critical aspect for biomarker discovery in PPMI, where hidden genetic and environmental factors may influence observed relationships. FCI extends the standard Markov blanket concept by leveraging the Possible-D-SEP (PDS) set, which consists of variables that might influence the relationship between two given nodes, even if they are not directly connected. The proposed PFCI method is conceptually similar to \cite{loh2014high}, but they do not consider the presence of latent or selection variables in the study.


The rest of the paper is organized as follows. In \Cref{sec:meth}, we introduce the penalized FCI algorithm and describe some important concepts associated with the method. In \Cref{sec:simu}, we conduct extensive simulation under widely varying setups to showcase the robustness of the proposed causal discovery approach. \Cref{sec:theory} develops the asymptotic consistency result of this method and delineates all the assumptions necessary for the theoretical guarantee. In \Cref{sec:realdata}, we thoroughly describe the data, the pre-processing, the analysis and finally the findings. We were able to identify one plasma biomarker (Myocilin) along with some well-studied CSF biomarkers in the first Markov blanket layer and several biomarkers in the second layer. In the Supplement, we have provided some additional simulation results and the proof of the theorem. 

\section{Statistical Method}
\label{sec:meth}
Consider $\bm{X} = (X_1,\dots,X_p)^\mathrm{T} \in \mathbbm{R}^p$. Let \( X_j \) denote each observed variable in a given dataset, where \( j \in \{1, 2, \dots, p\} \) represents the indices of these variables. Suppose that $\bm\Sigma$ is nonsingular. Each \( X_j \) could represent a distinct biomarker, feature, or other observable quantity that captures relevant information about the system of interest, with the goal being to identify causal pathways and dependencies among these \( X_j \)'s. The conditional independence structure of this distribution is represented by a network $\mathcal{G} = (\bm{V},E)$, where $\bm{V} = \{1,\dots,p\}$ is the set of vertices or nodes in the graph and $E$ is the set of edges in $\bm{V} \times \bm{V}$. Here, we consider the number of nodes to increase with the sample size $n$, i.e. $p=p_n$, and $V= V_n$. However, along with observed variables, say $\bm{O}$, here we also consider the presence of latent variables $\bm{L}$ which are hidden, and selection variables $\bm{S}$ which determine whether a particular observation appears in the dataset, introducing potential selection bias. Thus $\bm{V} = \bm{O} \cup \bm{L} \cup \bm{S}.$ Given the presence of selection bias due to $\bm{S}$, $(O_1,\dots,O_p)$ do not follow their marginal distribution but rather their conditional distribution given $\bm{S}$. This leads to the representation $\bm{X} = (X_1,\dots,X_p) \sim (O_1|\bm{S},\dots,O_p|\bm{S}).$ The edge $(i,j) \in E$ only if $X_i$ conditionally dependent on $X_j$ given all remaining variables. Suppose that we have $n$ observations on the $p$-dimensional vector $ (X_1,\dots,X_p)^\mathrm{T}$ which are stored in the matrix $\bm{X} \in \mathbbm{R}^{n \times p}$. 

In the first stage of the two-stage PFCI algorithm, we use the neighborhood selection method to construct an initial undirected dependency structure among the variables. Neighborhood selection is a regression-based technique that identifies conditional dependencies by examining the sparse precision matrix among variables. Specifically, for each variable \( X_j \), a penalized regression model is fitted using the Lasso (Least Absolute Shrinkage and Selection Operator) as follows:

\[
\hat{\beta}_{j} = \arg \min_{\beta} \left\{ \frac{1}{2n}  (\bm{X}_j - \bm{X}_{-j} \bm\beta_{j})^2 + \lambda_n \sum_{k \neq j} |\beta_{jk}| \right\},
\]
where \( \lambda_n \) is a tuning parameter controlling the sparsity of the solution and $\bm\beta_j = (\beta_{jk}: k \neq j) \in \mathbbm{R}^{p - 1}$ is the vector of regression coefficients from regressing $X_j$ on the rest of the variables. The neighborhood of \( X_j \), defined as the set of nodes with nonzero coefficients \( \beta_{jk} \text{ for } k \neq j \), provides the variables that are conditionally dependent on \( X_j \). By performing this selection for each variable, a symmetric adjacency matrix is constructed, yielding an undirected graph that captures the dependencies among \( X_j \)'s without directed edges. This adjacency structure serves as input for the next stage, where directed causal relationships are inferred.

In the second stage, we apply the FCI algorithm to convert the undirected skeleton into a Partial Ancestral Graph (PAG) while considering the possibility of latent (unobserved) confounders. In presence of hidden confounders, every DAG can be uniquely transformed into a Maximal Ancestral Graph (MAG). In other words, a MAG represents a causal structure among the observed variables, accounting for latent variables and selection bias. Several DAGs can thus result in the same MAG. A MAG contains mixed edge types ($\rightarrow, \leftrightarrow, \mathrel{-\circ},  \mathrel{\circ\!\!-\!\!\circ}, \mathrel{\circ\!\! \rightarrow} $) to represent ambiguity in causal directionality due to latent confounders, where $\rightarrow$ represents a directed edge, $\leftrightarrow$ represents a bi-directed edge, and $\mathrel{-\circ},  \mathrel{\circ\!\!-\!\!\circ}, \mathrel{\circ\!\! \rightarrow}$ represent undirected, nondirected and partially directed edges respectively. A DAG only consists of directed edges, whereas a MAG cannot be further refined without additional assumptions or interventional data. A PAG represents a Markov equivalence class of MAGs, meaning all MAGs in the equivalence class encode the same conditional independence relationships. 

FCI operates in two main steps: conditional independence testing and orientation propagation, using orientation rules that account for the existence of hidden confounders. The pairs of variables deemed conditionally independent by the first step of this hybrid approach are still considered the same while applying the FCI algorithm to the discovered skeleton. Thus, we are essentially running FCI on a sparser subgraph, thereby making it computationally much faster. The FCI algorithm’s orientation rules extend to scenarios where latent confounding may be present, incorporating bi-directed edges \( (X_i \leftrightarrow X_j) \) to indicate a shared unobserved confounder influencing both \( X_i \) and \( X_j \).

The FCI algorithm assumes a faithfulness condition (\Cref{assum:faith}), which is essential for reliable inference. However, by pruning edges in the graph with wrong or unreasonably high penalty, we may risk omitting real conditional dependencies that may otherwise appear in a fully faithful graph. In practical terms, faithfulness is generally maintained as long as the sparsity threshold is chosen carefully. Techniques like cross-validation can help ensure that important dependencies are retained, balancing faithfulness with interpretability by preserving core relationships without excessive sparsity. Moreover, \Cref{thm:main_thm} in Section \ref{sec:theory} establishes that for the correct theoretical choice of the penalty parameter, the PAG produced by PFCI is consistent for the true underlying DAG, and hence satisfies the faithfulness assumption. This choice of the tuning parameter leads to more time-efficient results compared to data-driven cross-validation techniques. This approach, integrating neighborhood selection and FCI, thus yields a causal DAG with directed edges that reflect potential causal relationships under latent confounding.

\section{Simulation Study}
\label{sec:simu}
In this section, we conduct two different simulation studies. In \Cref{sec:simu1}, we consider a general sparse DAG and try to see how the proposed method performs in terms of identifying the causally influenced nodes. In \Cref{sec:simu2}, we design a DAG that perhaps has more resemblance to real biomarker connections in neurodegerative diseases. We try to mimic the biomarker community structure by creating groups of highly linked nodes, out of which some are more causal than the others.

\subsection{Simulation Study 1}\label{sec:simu1}

In our simulation, we generate a DAG with $p$ nodes, ensuring sparsity constraints consistent with Assumption \ref{assum:sparsity}. We varied the node sizes ($p$) from 100 to 1000 while maintaining a consistent sparsity structure. In order to comply with the sparsity assumption required for the theoretical proof, we introduced directed edges from any node with probability $\pi = $0.015. Consequently, $c_3$ and $c_4$ from \Cref{assum:sparsity} should be so chosen that they follow the conditions $p\cdot \pi \approx n^{c_3}$ and $p^2 \cdot \pi \approx c_4$ respectively. Then for $\pi = 0.015$ and $p = \{100,200,\dots,1000\},$ we get $c_3 \approx \{0.09, 0.24, 0.33, 0.39, 0.44, 0.48, 0.52, 0.54, 0.56, 0.59\}$ and $c_4 = \{0.0225, 0.045, 0.0675, 0.09, 0.1125, 0.135, 0.1575,0.18, 0.2025, 0.225\}$. Our constructed DAG, thus, abides by the sparsity assumptions required to achieve consistency of the estimated causal structure. We generated data from multivariate normal distribution based on the aforementioned DAG structure with a sample size of 100. Given the large number of nodes relative to the sample size, this setup represents a high-dimensional regime, where the number of variables far exceeds the number of observations. To implement the FCI and RFCI algorithms, we used the \texttt{pcalg} R package \citep{kalisch2012causal}.

In order to assess whether the proposed method is able to recover the true causal DAG structure, we report the Structural Hamming Distance (SHD), The Matthew's Correlation Coefficient (MCC) and the F1-score. SHD is a measure of the difference between the estimated and true graph structures, calculated as the total number of edge additions, deletions, and reversals needed to transform the estimated graph into the true graph. Lower SHD values indicate closer alignment to the true causal structure, with zero representing perfect reconstruction. F1-score, on the other hand, provides a balance between precision and recall in detecting the causal edges, where higher values indicate better identification of the undirected edges. Suppose $TP$ is the number of true positives, $FP$ is the false positives denoting the count of falsely selected edges, $TN$ (True Negative) is the count of independence relationships correctly identified by the model, and $FN$ (False negative) is the number of causal edges mistakenly left out by the model.
Then the F1-score is mathematically defined by $F_1 = ({2*\textnormal{Precision}*\textnormal{Recall}})/({\textnormal{Precision}+\textnormal{Recall}})$ where $\textnormal{Precision} = TP/(TP + FP)$ and $\textnormal{Recall} = TP/(TP + FN)$. Finally, MCC is defined as $\text{MCC} = ({\text{TN}\times \text{TP} - \text{FN}\times \text{FP}})({\sqrt{(\text{TP+FP})(\text{TP+FN})(\text{TN+FP})(\text{TN+FN})}})$.

As shown in \Cref{tab:my-table_general_DAG}, the proposed PFCI method performs at par or better than RFCI in terms of SHD, F1-score and MCC while achieving significantly better computational efficiency. The improvement in structure recovery is more pronounced for higher dimensions. Moreover, from the table, it can be safely claimed that our method is approximately 3 times faster than the state-of-the-art RFCI method, while producing better estimation accuracy. 
\begin{table}[htbp]
\centering
\resizebox{0.9\textwidth}{!}{%
\begin{tabular}{c|cc|cc|cc|cc}
\hline
\multirow{2}{*}{Setup} & \multicolumn{2}{c|}{SHD} & \multicolumn{2}{c|}{F1-score} & \multicolumn{2}{c|}{MCC} & \multicolumn{2}{c}{Run-time} \\ \cline{2-9} 
 & PFCI & RFCI & PFCI & RFCI & PFCI & RFCI & PFCI & RFCI \\ \hline
$p = 100$ & 80.70 (5.19) & 79.70 (6.53) & 0.74 (0.02) & 0.76 (0.02) & 0.75 (0.01) & 0.76 (0.01) & 0.13 (0.01) & 0.22 (0.02) \\
$p = 200$ & 207.15 (10.21) & 208.70 (10.66) & 0.76 (0.02) & 0.75 (0.01) & 0.75 (0.02) & 0.75 (0.02) & 0.47 (0.04) & 1.01 (0.09) \\
$p = 300$ & 354.35  (11.82) & 356.35 (11.78) & 0.74 (0.03) & 0.74 (0.03) & 0.75 (0.03) & 0.75 (0.02) & 0.99 (0.05) & 2.51 (0.13) \\
$p = 400$ & 520.20 (11.5) & 541.2 (14.8) & 0.79 (0.02) & 0.77 (0.02) & 0.79 (0.02) & 0.77 (0.02) & 1.89 (0.13) & 5.37 (0.30) \\
$p = 500$ & 728.00 (17.95) & 782.85 (18.9) & 0.78 (0.02) & 0.76 (0.02) & 0.78 (0.02) & 0.75 (0.02) & 3.31 (0.21) & 9.25 (0.5) \\
$p = 600$ & 966.86 (29.05) & 1047.07 (27.07) & 0.75 (0.02) & 0.73 (0.01) & 0.76 (0.02) & 0.73 (0.01) & 5.79 (0.27) & 15.24 (1.1) \\
$p = 700$ & 1272.35 (26.1) & 1374.45 (28.6) & 0.71 (0.02) & 0.70 (0.01) & 0.73 (0.01) & 0.71 (0.01) & 9.33 (0.42) & 23.18 (1.31) \\
$p = 800$ & 1507.43 (29.68) & 1615.86 (31.24) & 0.72 (0.03) & 0.69 (0.04) & 0.71 (0.04) & 0.68 (0.04) & 15.25 (1.25) & 37.04 (2.89) \\
$p = 900$ & 2075.10 (39.97) & 2190.05 (47.07) & 0.70 (0.02) & 0.67 (0.02) & 0.70 (0.01) & 0.66 (0.02) & 21.41 (2.38) & 55.33 (5.51) \\
$p = 1000$ & 2571.45 (47.17) & 2687.85 (40.21) & 0.69 (0.01) & 0.64 (0.01) & 0.67 (0.01) & 0.64 (0.01) & 39.52 (2.40) & 120.07 (10.61) \\ \hline
\end{tabular}%
}
\caption{Comparison of SHD, F1-score, MCC and running times (in seconds) of PFCI and RFCI for a general sparse DAG.}
\label{tab:my-table_general_DAG}
\end{table}
In order to check the performance robustness of the method when the underlying data generation process is non-normal, we carry out another simulation study. Here we use the $t$-distribution with 4 degrees of freedom. In \Cref{tab:my-table_general_DAG_non-normal}, we provide the performance metrics corresponding to the two competing methods. We observe similar patterns in the F1-score, MCC as well as in the SHD under similar time-complexity even under the non-normal case. Although SHD is similar in both Tables \ref{tab:my-table_general_DAG} and \ref{tab:my-table_general_DAG_non-normal}, F1-scores and MCC suffer a little for both methods in the latter case. However, PFCI still mostly leads in terms of node identification and edge orientation.

\begin{table}[]
\centering
\resizebox{0.9\textwidth}{!}{%
\begin{tabular}{c|cc|cc|cc|cc}
\hline
\multirow{2}{*}{Setup} & \multicolumn{2}{c|}{SHD} & \multicolumn{2}{c|}{F1-score} & \multicolumn{2}{c|}{MCC} & \multicolumn{2}{c}{Run-time} \\ \cline{2-9} 
 & PFCI & RFCI & PFCI & RFCI & PFCI & RFCI & PFCI & RFCI \\ \hline
$p = 100$ & 82.90 (6.23) & 83.60 (6.35) & 0.66 (0.03) & 0.67 (0.02) & 0.64 (0.02) & 0.64 (0.02) & 0.18 (0.01) & 0.24 (0.02) \\
$p = 200$ & 210.40 (9.12) & 209.00 (10.19) & 0.69 (0.02) & 0.69 (0.01) & 0.68 (0.02) & 0.68 (0.03) & 0.65 (0.03) & 1.06 (0.06) \\
$p = 300$ & 357.60 (11.22) & 361.20 (13.12) & 0.73 (0.01) & 0.73 (0.01) & 0.74 (0.01) & 0.74 (0.02) & 1.43 (0.10) & 2.62 (0.17) \\
$p = 400$ & 530.70 (13.63) & 556.40 (15.73) & 0.76 (0.01) & 0.74 (0.02) & 0.76 (0.02) & 0.75 (0.01) & 2.71 (0.18) & 5.30 (0.31) \\
$p = 500$ & 734.80 (19.41) & 781.80 (18.70) & 0.77 (0.01) & 0.75 (0.01) & 0.77 (0.01) & 0.75 (0.02) & 4.44 (0.27) & 8.23 (0.52) \\
$p = 600$ & 977.35 (22.25) & 1050.5 (23.87) & 0.74 (0.02) & 0.73 (0.02) & 0.75 (0.01) & 0.73 (0.02) & 7.61 (0.43) & 13.47 (1.11) \\
$p = 700$ & 1298.40 (27.28) & 1393.20 (27.79) & 0.68 (0.02) & 0.69 (0.01) & 0.70 (0.02) & 0.69 (0.01) & 13.19 (1.12) & 21.85 (2.34) \\
$p = 800$ & 1675.60 (29.92) & 1783.40 (30.12) & 0.66 (0.02) & 0.64 (0.01) & 0.65 (0.03) & 0.64 (0.03) & 24.09 (2.22) & 39.44 (2.94) \\
$p = 900$ & 2013.52 (42.31) & 2164.29 (41.35) & 0.66 (0.02) & 0.65 (0.02) & 0.66 (0.02) & 0.64 (0.02) & 33.32 (3.02) & 63.67 (5.52) \\
$p = 1000$  & 2571.40 (51.45) & 2687.90 (46.66) & 0.63 (0.01) & 0.63 (0.02) & 0.64 (0.01) & 0.63 (0.01) & 48.57 (3.31) & 121.98 (9.58) \\ \hline
\end{tabular}%
}
\caption{Comparison of SHD, F1-score, MCC and running times (in seconds) of PFCI and RFCI for a general sparse DAG under non-normal data generation process.}
\label{tab:my-table_general_DAG_non-normal}
\end{table}

\subsection{Simulation Study 2}\label{sec:simu2}
We conduct a second simulation, where we are interested in a specific structure of DAG. Consider a directed graph where nodes are grouped based on correlation, with sparse inter-group connections and denser within-group connections. Moreover, we have a response variable that depends only on a few specific groups, thereby introducing a hybrid structure combining local dense clusters with global sparsity. Such sparsity patterns are common in applications like brain connectivity studies, where regions of the brain may exhibit dense intra-regional connections, with sparser connections between regions, and certain regions may be more influential in driving behavioral or cognitive outcomes. However, directed graphical models with group-specific dependencies for causal inference, particularly incorporating a response variable that depends on only a subset of these groups, remain relatively underexplored, signaling a potential area for further research.


In Parkinson’s disease research, data such as those from the PPMI might include groups of blood-based and CSF biomarkers. Biomarkers within a single biological process, for example, inflammatory or metabolic pathways, often exhibit high correlations, forming natural groups within the data. A variable of interest, such as disease progression or response to treatment, could influence some, but not all, of these groups. Understanding which specific clusters of biomarkers are causally impacted by disease status has significant implications for targeted therapeutic interventions and for understanding the biological mechanisms underlying disease progression. This grouping phenomenon is commonly observed in neurodegenerative diseases as discussed in \cite{calderone2016comparing, vrahatis2020detecting, han2022biomarkers}.

\begin{figure}
    \centering
    \includegraphics[width=\linewidth]{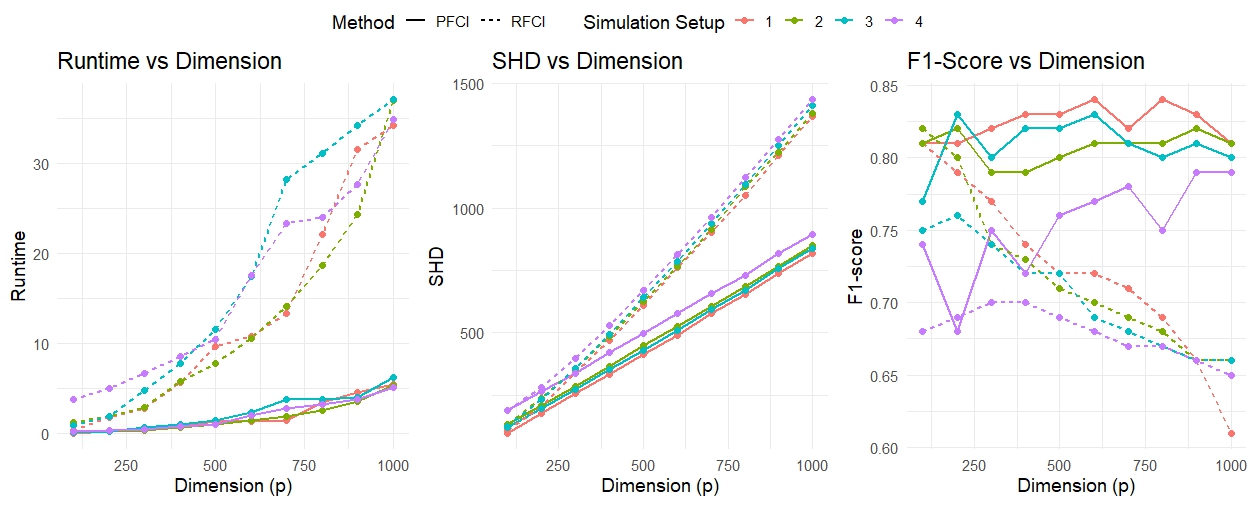}
    \caption{Comparison of PFCI \textit{(solid lines)} and RFCI \textit{(dashed lines)} across four simulation setups \textit{(different colors)}. The plots show \textit{(left)} run-time in seconds, \textit{(middle)} SHD, and \textit{(right)} F1-score, all as functions of the dimension \( p \).}
    \label{fig:Simulation2}
\end{figure}

We simulate data from a DAG representing a network of variables organized into groups, with a root variable \( Y \) as the primary causal factor. Each group contains \( K \) correlated variables, with some groups directly influenced by \( Y \) while others remain irrelevant. Sparse between-group connections exist but do not imply causal links beyond the first or, at most the second Markov blanket layer. To evaluate our method’s scalability, we vary group size \( K \) $(5, 10)$ and total nodes \( p \) (100 to 1000 in steps of 100), adjusting the number of groups accordingly (e.g., \( p = 100, K = 5 \) implies 20 groups). We also vary sparsity effects by selecting \( s = 5,10 \) causal groups where \( Y \) influences most variables. This setting ensures a realistic yet tractable structure, demonstrating our method’s efficiency compared to RFCI as \( p \) grows. 
We have used four simulation set-ups (1: $K=5, s=5$; 2: $K=10, s=5$; 3: $K=5, s=10$; 4: $K=10,s=10$). Figure \ref{fig:Simulation2} provides a visual representation of runtime, SHD and F1-Score in all four set-ups. Our method demonstrates huge computational gain for higher dimensions, that is, for higher values of $p$.
Additionally, in all four set-ups, we observe that the SHD of PFCI is uniformly lower than that of RFCI, meaning that fewer alterations are required by the PFCI to get back to the true DAG as compared to RFCI. We also see that $F_1$ is mostly higher for the PFCI method in all four settings. The exact numbers along with their standard deviations are provided in Tables S1-S4 in the Supplement.

\section{Theoretical Results}
\label{sec:theory}
We first introduce some notations before stating the main theorem. Let $\mathcal{D}_0$ be the true underlying DAG that we are trying to capture. We assume the true network structure to follow some sparisty assumption stated later. We denote the skeleton of $\mathcal{D}_0$ by $\mathcal{U}_0$. Given data $\bm{X}$, let $\bm{\Omega}$ be the covariance matrix on which we apply the neighborhood selection technique. Let $\hat{\mathcal{U}}^{\mathrm{NS}}$ denote the sparse estimator obtained by the Neighborhood Selection (NS) procedure. We define $\hat{\mathcal{D}}^\mathrm{NS}$ as the underlying DAG of $\hat{\mathcal{U}}^{\mathrm{NS}}$ obtained by imposing directionalities to the selected edges. Finally, let $\hat{\mathcal{D}}^\mathrm{PFCI}$ be the proposed estimator obtained by performing the FCI operation only on the skeleton network chosen by the neighborhood selection method. We aim to show that the proposed penalized FCI estimator $\hat{\mathcal{D}}^\mathrm{PFCI}$ asymptotically approaches the true DAG $\mathcal{D}_0$. 

Next, we discuss some important needed by the FCI algorithm to orient the undirected edges. A set of nodes $\bm{V}_1$ is said to be \textit{d-separated} from a set $\bm{V}_2$ given another set of verices $\bm{Z}$ in a DAG if every path between any node in $\bm{V}_1$ and any node in $\bm{V}_2$ is blocked by $\bm{Z}$. A path can be blocked in three ways: if, for $M \in \bm{Z}$, $v_1 \in \bm{V}_1$ and $v_2 \in \bm{V}_2,$ there exists (1) a chain $v_1 \to M \to v_2$, (2) a fork $v_1 \leftarrow M \to v_2$,
and if, for $M$ and any of its descendants not included in $\bm{Z}$, there exists (3) a collider $v_1 \to M \leftarrow v_2$. An \textit{m-separation}, on the other hand, is a generalization of d-separation in MAGs, which allow for latent variables and undirected or bidirected edges. If two vertices $v_1$ and $v_2$ are d-separated in a DAG by $\bm{Z} \subseteq \bm{V} \setminus \{v_1,v_2\},$ then $v_1 \perp v_2 | \bm{Z}$ in any distributuion that factorizes according to the DAG. The distribution will be said to be \textit{faithful} to the DAG if the conditional independence relation in the distribution are exactly the same as those obtained from the DAG using d-separation. m-separation, thus, additionally accounts for hidden confounders. For example, in $v_1 \leftarrow U \to v_2$, if $U$ is unobserved, a MAG wil show $v_1 \leftrightarrow v_2$. d-separation would have failed in such a case, but m-separation will be able to detect the path blockage. Therefore, if $v_1$ and $v_2$ are m-separated in a MAG by $\bm{Z}$, then $v_1$ and $v_2$ are d-separated in the corresponding DAG by $\bm{Z} \cup \bm{S}$ and consequently $v_1 \perp v_2 | \bm{Z}  \cup \{\bm{S} = \bm{s}\}$. Lastly, we define an FCI-PAG ($C$) of an underlying DAG \( \mathcal{G} \) over a vertex set partitioned into observed variables \( \mathbf{O} \), latent variables \( \mathbf{L} \), and selection variables \( \mathbf{S} \) satisfies four conditions:

\begin{enumerate}
    \item \textit{Conditional Independence Representation}: If no edge exists between two observed variables \( v_i \) and \( v_j \) in \( C \), there must be some subset \( \mathbf{Y} \subseteq \mathbf{O} \setminus \{v_i, v_j\} \) such that \( v_i \) and \( v_j \) are conditionally independent given \( \mathbf{Y} \cup \mathbf{S} \).
    \item \textit{Edge Necessity}: If an edge exists between \( v_i \) and \( v_j \) in \( C \), then they are not conditionally independent given any subset \( \mathbf{Y} \).
    \item \textit{Arrowhead Interpretation}: If an edge between \( v_i \) and \( v_j \) in \( C \) has an arrowhead at \( v_j \), then \( v_j \) is not an ancestor of \( v_i \) in \( \mathcal{G} \), even when including selection variables.
    \item \textit{Tail Interpretation}: If an edge between \( v_i \) and \( v_j \) in \( C \) has a tail at \( v_j \), then \( v_j \) must be an ancestor of \( v_i \).
\end{enumerate}
We now state some essential assumptions. The validity of these assumptions for the PPMI biomarker data are discussed later in \Cref{sec:subjects}. A general discussion of the assumptions may be found in the Supplement.

\begin{assumption}[High-dimensionality]\label{assum:high-dim}
    For $c_1>0$, we have $p = \mathcal{O}(n^{c_1})$. 
\end{assumption}

\begin{assumption}[Nonsingularity]\label{assum:nonsingularity}
    For all $j \in V$, we have $\textnormal{var}(X_j) = 1$ and there exists $c_2^2 > 0$, such that for all $n \in \mathbb{N}$ and $j \in V$, $\textnormal{var}(X_j|\bm{X}_{V\setminus \{j\}}) \geq c_2^2.$
\end{assumption}

\begin{assumption}[Sparsity]\label{assum:sparsity} We need to impose two sparsity constraints on the undirected network. The first one restricts the maximum number of edges for each node in the graph and the second restricts the number of mutual neighbors of two neighboring nodes. Suppose the neighborhood $n(j)$ of a node $j \in V$ is the smallest subset of $V \setminus \{j\}$ such that, given all variables in the neighbborhood, $X_j$ is conditionally independent of all remaining variables. 
        \begin{enumerate}
            \item There exists $c_3 \in [0,1)$ such that $\max_{j \in V} \lvert{n(j)}\rvert$ $ = \mathcal{O}(n^{c_3})$.
            \item There exists some $c_4 < \infty$ such that $\max_{j \in n(i)}|n(j) \cap n(i)| \leq c_4$.
        \end{enumerate}
\end{assumption}

\begin{assumption}[Possible-D-sep]\label{assum:possible_Dsep}
    The maximum size of the possible-D-sep sets allowed in the FCI algorithm is denoted by $r_n$, and we assume $r_n = \mathcal{O}(n^{1-c_5})$ for $0 < c_5 \leq 1$
\end{assumption}

\begin{assumption}[Partial Correlations]\label{assum:par-corr}
    We define $\pi_{ij}$ as the partial correlation between $X_i$ and $X_j$, which refers to the correlation after removing the linear effects from all remaining variables $\{X_k : k \in V \setminus \{i,j\}\}$. There exists $c_6 > 0$ and $1 \geq c_7 > \max\{c_3, 1 - c_5\}$ such that for every $(i,j) \in E$, we have $\underset{i,j\lvert \pi_{ij} \neq 0}{\inf}\{|\pi_{ij}|\} \geq c_6 n^{-(1 - c_7)/2} \textnormal{ and } \underset{i \neq j}{\sup}\{|\pi_{ij}|\} \leq c_8 < 1.$
\end{assumption}

\begin{assumption}[Neighborhood Stability]\label{assum:neigh_stability} We first define $\bm\beta^{j,\mathcal{A}} = \argmin_{b: \forall k \in \mathcal{A}, b_k \neq 0} \mathbbm{E}\Big(X_j - \sum_{k \in V} b_k X_k \Big)^2 \in \mathbbm{R}^{|\mathcal{A}|}$ and consequently $S_j(l) = \sum_{k \in n(j)} \textnormal{sign}(\beta_{k}^{j,n(j)})\beta_{k}^{l,n(j)}.$ Then, there exists some $c_9 < 1$ so that for all $(j,l) \in V$ with $l \notin n(j),$ we assume $|S_j(l)| < c_9$.
\end{assumption}

\begin{assumption}[Faithfulness]\label{assum:faith}
    The distribution of $\bm{X}$ is faithful to the underlying causal MAG for all $n$.
\end{assumption}

\begin{assumption}[Distribution]\label{assum:MVN}
    The distribution of $\bm{X}$ is multivariate normal.
\end{assumption}

We show next that, under the above stated assumptions, $\hat{\mathcal{D}}^\mathrm{PFCI}$ matches $\mathcal{D}_0$ with high limiting probability, that is,
the PAG produced by implementing FCI over the sparsified undirected network obtained by the neighborhood selection approach is capable of recovering the true underlying
PAG. The proof of Theorem \ref{thm:main_thm} is given in the Supplement.

\begin{theorem}\label{thm:main_thm}
    Under Assumptions 1-8, $p = \mathcal{O}(\textnormal{exp}(m_4 n^{c_7} - \log n))$ and $\lambda_n \asymp m_3 n^{-(1-m_2)/2}$, we can show that $\P(\hat{\mathcal{D}}^\mathrm{PFCI} = \mathcal{D}_0) \to 1$ as $n \to \infty.$
\end{theorem}

\section{Real Data Analysis}
\label{sec:realdata}

In this section, we analyze proteomic data from the Project 9000 of PPMI and Olink platforms, incorporating detailed patient information from the study subjects subsection. 

\subsection{Project 9000}
The Parkinson's Progression Markers Initiative (PPMI) is a landmark study aimed at identifying biomarkers for Parkinson's disease (PD) progression. This initiative collects extensive clinical, imaging, and biological data from participants, including CSF and blood samples, to elucidate the pathophysiological mechanisms underlying PD. The PPMI cohort is predominantly composed of individuals diagnosed with early-stage PD, along with healthy controls, allowing for a comprehensive analysis of disease progression and the identification of potential biomarkers that could aid in diagnosis and treatment \citep{stewart2019impact}. The integration of genetic, clinical, and proteomic data within the PPMI framework enhances the understanding of PD's heterogeneous nature, facilitating the identification of biomarkers that correlate with clinical outcomes and disease progression \citep{stewart2019impact}. Project 9000, a significant component of the PPMI, leverages advanced proteomic technologies to analyze a vast array of proteins in biological samples. Project 9000 contains bridged results of Projects 196 and 222, both of which analyzed cerebrospinal fluid and plasma for a limited number of participants. One of the key platforms utilized in this project is the Olink Explore 1536 targeted proteomics (\url{https://www.olink.com}), which we shall discuss next.

\subsection{Olink}
The Olink Explorer employs Proximity Extension Assay (PEA) technology to measure proteins in biological samples like plasma, serum, CSF and to enable high-throughput analysis of protein biomarkers from minimal sample volumes (as low as 1 $\mu L$) \citep{katz2022proteomic}. The Olink Explore platform is particularly advantageous due to its ability to measure a large number of proteins simultaneously, with the latest panels assessing up to 1536 proteins \citep{navrazhina2022large}. This high multiplexing capability allows for the identification of unique protein signatures associated with various diseases, including PD, and provides insights into the underlying biological processes \citep{navrazhina2022large}. Furthermore, studies have shown that the Olink platform demonstrates superior performance in terms of assay specificity and sensitivity compared to traditional immunoassay methods, making it a valuable tool for biomarker discovery \citep{eldjarn2022large}.

The advantages of Olink Explore over existing biomarker platforms are underscored by its ability to detect low-abundance proteins that are often missed by other techniques. For instance, in a comparative study, Olink assays exhibited a significantly higher fraction of cis-acting protein quantitative trait loci (pQTLs), indicating that the platform effectively measures the intended proteins \citep{eldjarn2022large}. This capability is crucial for understanding the genetic underpinnings of diseases and their associated biomarkers. Additionally, the Olink platform has been utilized to uncover novel protein associations in various conditions, such as COVID-19, where it identified distinct protein signatures that correlate with disease severity \citep{wik2021proximity}. The integration of such findings into the PPMI framework could enhance the identification of biomarkers that predict disease progression and therapeutic responses in PD patients. 

PPMI Project 9000 follows rigorous quality control (QC) procedures to ensure reliability, accuracy, and reproducibility of the high-dimensional protein expression data obtained through Olink Explore 1536 panels. The QC steps help mitigate batch effects, ensure data consistency across samples, and remove unreliable measurements. Each Olink panel includes internal controls to assess assay performance and adjust for technical variations across plates and runs. Only proteins passing sample-level QC, batch-level QC and batch-effect correction criteria are retained for downstream analysis. After applying these QC measures, the final dataset contains high-confidence protein biomarkers with reduced batch effects and improved reproducibility, making it suitable for discovery and inference.

\subsection{Study Subjects}\label{sec:subjects}
The dataset utilized in this study comprises CSF and plasma biomarkers from 199 patients, encompassing a total of 2,924 biomarkers, with 1,462 derived from each biofluid. We first assess the validity of key model assumptions within the PPMI dataset. This evaluation ensures that our approach is appropriate for this case study and helps contextualize the interpretation of the results.

\rowcolors{2}{white}{gray!15} 
\begin{longtable}{p{3cm} p{12cm}} 
\toprule
\rowcolor{gray!20} 
\textbf{Assumptions} & \textbf{Justification for the PPMI Data} \\ 
\endfirsthead
\multicolumn{2}{c}%
{{\bfseries Table \thetable\ Continued from previous page}} \\
\hline
\rowcolor{gray!20} \textbf{Assumption} & \textbf{Justification for the PPMI} \\
\endhead

\midrule

\endfoot

\bottomrule
\endlastfoot

Assumption \ref{assum:high-dim} & We have $p = 2924$ biomarkers and only $n = 199$ patients. \\ 
Assumption \ref{assum:nonsingularity} & The data has been properly normalized to meet this assumption. \\ 
Assumption \ref{assum:sparsity}(i) & Essential with 2,924 biomarkers but only 199 patients, to ensure identifiability and avoid overfitting. \\ 
Assumption \ref{assum:sparsity}(ii) & Ensures that highly connected subgraphs (such as cliques) do not form excessively in the biomarker network.  
Studies constructing biomarker networks in neurodegenerative diseases (e.g., Alzheimer's Disease) often  
find that key biomarkers cluster into small, functionally related groups.  
For example, CSF biomarkers of tau, amyloid, and neuroinflammation tend to interact within small subnetworks  
rather than forming a large fully connected group. \\ 
Assumption \ref{assum:possible_Dsep} & Causal relationships in biomarker networks tend to be localized, meaning that most biomarkers interact in  
small functional modules rather than forming highly connected, large-scale dependencies.  
This assumption ensures that the possible-D-sep set remains biologically meaningful by keeping conditioning sets small,
 preventing spurious causal links due to large, irrelevant conditioning sets. \\ 
Assumption \ref{assum:par-corr} & In biomarker networks for neurodegenerative diseases, the lower bound on partial correlations prevents false  
discoveries by ensuring that detected edges are biologically meaningful, while the upper bound prevents certain  
biomarkers from dominating the network, ensuring stable causal inference. \\ 
Assumption \ref{assum:neigh_stability} & Ensures that biomarkers outside of a direct pathway do not have undue influence on a node, aligning with  
biological modularity. \\ 
Assumption \ref{assum:faith} & Ensures that measured variables (biomarkers) have meaningful dependencies rather than coincidental ones.  
PPMI collects multiple biological, clinical, and imaging biomarkers, making it less likely that independencies  
arise purely due to sampling variability. \\ 
Assumption \ref{assum:MVN} & Many biomarkers, which are continuous variables, exhibit approximately Gaussian behavior when appropriately  
normalized. Moreover, simulation results indicate that violations  
of this assumption will not hugely impact the correctness of the findings. \\ 

\end{longtable}%

\begin{figure}[htbp]
    \centering
    \includegraphics[width=\textwidth]{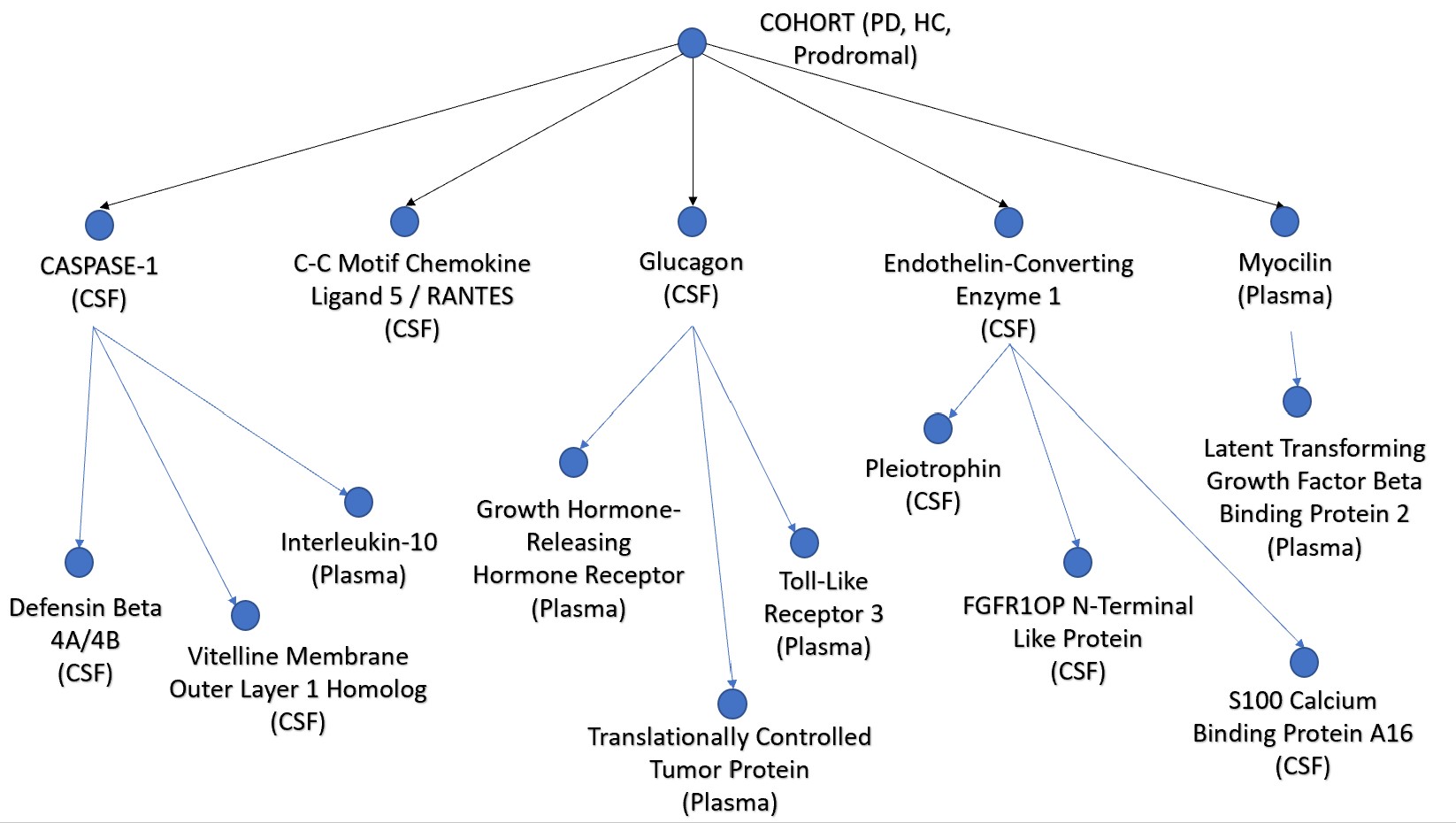}
    \caption{The Markov blanket extracted from the causal network, illustrating the biomarkers with a direct or indirect causal relationship with disease classification (PD, healthy controls, and prodromal cases). The first layer consists of biomarkers directly connected to the cohort node, representing key inflammatory, metabolic, and neurovascular processes. The second layer comprises downstream biomarkers influenced by the first-layer connections, potentially acting as modulators or effectors of disease progression. CSF and plasma biomarkers are distinguished based on their source, highlighting their respective roles in Parkinson’s disease pathophysiology.}
    \label{fig:blanket}
\end{figure}

Initially, we construct a causal network using the algorithm proposed in this paper. Protein expression data for each biomarker and for all subjects are presented in Normalized Protein eXpression (NPX) units, which are log2-transformed and normalized. NPX values allow for comparative analysis across samples but require proper normalization for cross-study comparisons. Subsequently, we extract the Markov blanket surrounding the cohort node to identify biomarkers that exhibit a causal relationship with disease progression, as illustrated in Figure \ref{fig:blanket}.

\subsection{Results}\label{sec:results}
The first layer of the Markov blanket consists of biomarkers directly causally associated with the disease classification (PD, healthy controls, and prodromal cases). These proteins represent key inflammatory, metabolic, and neurovascular pathways, which have been widely implicated in PD. Caspase-1, a crucial component of the inflammasome, is known to drive neuroinflammation in PD through its activation of interleukin-1$\beta$ (IL-1$\beta$), leading to dopaminergic neuron death and $\alpha$-synuclein aggregation, which are hallmarks of the disease \citep{wang2016caspase}. Studies have shown that caspase-1 inhibitors can reduce neuroinflammation and delay disease progression in PD models \citep{gordon2018inflammasome, gao2020through}. Similarly, C-C motif chemokine ligand 5 (CCL5/RANTES) plays a role in T-cell migration and microglial activation, which are key drivers of chronic neuroinflammation in PD \citep{zhao2023ccl5}. Elevated CCL5 levels have been observed in PD patients, correlating with increased neurodegeneration and disease severity \citep{tang2014correlation, sahin2016case}. Another key first-layer protein, Glucagon, is associated with metabolic regulation and insulin signaling, both of which are emerging as important contributors to PD pathogenesis \citep{athauda2016glucagon, reich2022neuroprotective}. Glucagon dysfunction has been linked to neurodegenerative processes through mitochondrial impairment and oxidative stress, though direct associations with PD are still under investigation \citep{mehta2021deciphering}. Meanwhile, Endothelin-Converting Enzyme 1 (ECE1), known for its role in neuropeptide processing, has been implicated in neurovascular dysfunction, which can contribute to blood-brain barrier breakdown and neuronal death in PD \citep{zlokovic2008blood, chen2018stress}. Finally, Myocilin, though primarily studied in glaucoma and extracellular matrix regulation, has been linked to structural protein integrity in the nervous system, with potential implications for neuronal resilience and PD susceptibility \citep{saccuzzo2023myocilin, carbone2009overexpression}. Notably, Myocilin is the only Plasma biomarker in the first level of Markov blanket, all the others being CSF biomarkers.

The second layer of the Markov blanket consists of biomarkers that are influenced by the first layer, suggesting that they may serve as modulators or downstream effectors of neurodegenerative processes in PD. Defensin Beta 4A/4B, a downstream effector of Caspase-1, is an antimicrobial peptide that has been shown to modulate innate immune responses, and its connection to inflammasome activation suggests a potential role in PD-associated neuroinflammation \citep{navrazhina2022large}. Similarly, Vitelline Membrane Outer Layer 1 Homolog (VMO1) has been implicated in cell adhesion and structural maintenance, though its specific role in PD remains underexplored \citep{sagvekar2018functional}. Interleukin-10 (IL-10), which is downstream of Caspase-1, is a potent anti-inflammatory cytokine, and research has demonstrated that enhancing IL-10 signaling in PD models can reduce microglial activation and prevent dopaminergic neuron loss \cite{su2022targeting}. This suggests a protective mechanism counteracting the effects of chronic inflammation in PD. Studies have also indicated elevated Interleukin levels in Alzheimer's and Parkinson's Disease patients \citep{brodacki2008serum, blum1995interleukin, rentzos2009circulating}. Additionally, Growth Hormone-Releasing Hormone Receptor (GHRHR), which is downstream of Glucagon, has been linked to neuroprotection through its role in metabolic regulation and neuronal growth, potentially influencing dopaminergic survival in PD \citep{friess2001increased}. Meanwhile, Toll-Like Receptor 3 (TLR3), which is also downstream of Glucagon, has been shown to promote $\alpha$-synuclein aggregation and trigger inflammatory cascades, further linking metabolic dysfunction to immune activation in PD \citep{heidari2022role, kouli2019toll}. Translationally Controlled Tumor Protein (TPT1) has been implicated in cell survival and apoptosis, and though its direct role in PD is unclear, it may influence neuronal stress responses \citep{miao2024multifaceted}.

Further within the second layer, several proteins associated with neurotrophic signaling and extracellular matrix regulation are evident. Pleiotrophin, which is downstream of ECE1, is a growth factor known to support neuronal survival, and studies suggest that it may have compensatory or neuroprotective functions in neurodegenerative diseases \citep{fernandez2017pleiotrophin, hida2007pleiotrophin}. FGFR1OP N-Terminal Like Protein, also linked to ECE1, is involved in cytoskeletal organization, and while it has not been widely studied in PD, its function may contribute to neuronal stability \citep{melillo2015ret}. Additionally, S100 Calcium Binding Protein A16, downstream of FGFR1OP, is part of the S100 protein family, which has been linked to neuroinflammatory and neurodegenerative processes, including glial activation in PD \citep{figura2021proteomic, goswami2023pharmacological}. Lastly, Latent Transforming Growth Factor Beta Binding Protein 2 (LTBP2), connected to Myocilin, is involved in extracellular matrix remodeling and TGF-$\beta$ signaling, both of which play roles in blood-brain barrier integrity and neurovascular health, making it a possible modulator of neurodegeneration in PD \citep{tsuruga2012latent}.

This Markov blanket highlights a hierarchical structure of interactions, where first-layer biomarkers serve as primary disease mediators, while second-layer biomarkers represent downstream regulatory pathways that may modulate disease progression. The interaction between metabolic dysfunction, neuroinflammation, and extracellular matrix remodeling in PD is evident, suggesting that targeting both primary and secondary biomarkers could be a potential therapeutic strategy. Future studies should focus on validating these interactions in larger cohorts and exploring their mechanistic roles in PD pathology.

\section{Acknowledgement}\label{sec:acknow}

Parkinson's Data used in the preparation of this article (Section \ref{sec:realdata}) were obtained from the Parkinson's Progression
Markers Initiative (PPMI) database (\href{www.ppmi-info.org/access-data-specimens/download-data}{www.ppmi-info.org/access-data-specimens/download-data}).
For up-to-date information on the study, visit \href{www.ppmi-info.org}{www.ppmi-info.org}.

\bibliographystyle{apalike}
\bibliography{Bibliography-MM-MC}

\end{document}